\documentclass[twocolumn,showpacs,preprintnumbers,amsmath,amssymb]{revtex4}
\usepackage{graphicx}
\usepackage{dcolumn}
\usepackage{bm}
\def\bea {\begin{eqnarray}}
\def\eea {\end{eqnarray}}

\def\be {\begin{equation}}
\def\ee {\end{equation}}

\begin{document}
\title{Indication of transverse radial flow in high-multiplicity proton-proton collisions at the Large Hadron Collider}
\author{Premomoy Ghosh$^1$\footnote{Corresponding author: prem@vecc.gov.in},  Sanjib Muhuri $^1$, Jajati K. Nayak$^1$ and Raghava Varma$^2$ }
\address{$^1$Variable Energy Cyclotron Centre, Kolkata 700 064, India}                 
\address{$^2$Indian Institute of Technology Bombay, Mumbai 400 076, India}            
\date{\today}

\begin{abstract}
We analyze the measured spectra of $\pi^\pm$, $K^\pm$, 
$p$($\bar p$) in $pp$ collisions at $\sqrt {s}$ = 0.9, 2.76 and 7 TeV, in the light of blast-wave model to extract the transverse radial 
flow velocity and kinetic temperature at freeze-out for the system formed in $pp$ collisions.  The dependency of the blast-wave 
parameters on average charged particle multiplicity of event sample or the `centrality' of collisions has been studied and compared 
with results of similar analysis in nucleus-nucleus ($AA$) and proton-nucleus ($pA$) collisions. We analyze the spectra of 
$K_{s}^0$, $\Lambda$($\bar \Lambda$) and $\Xi^-$ also to see the dependence of blast-wave description on the species of produced
particles. Within the framework of the blast-wave model, the study reveals indication of collective behavior for high-multiplicity 
events in $pp$ collisions at LHC. Strong transverse radial flow in high multiplicity $pp$ collisions and its comparison with that in  $pA$ and 
$AA$ collisions match with predictions from a very recent theoretical work [Shuryak and Zahed 2013 arXiv:hep-ph/1301.4470] that addresses the conditions for 
applicability of hydrodynamics in $pp$ and $pA$ collisions.
\end{abstract}

\pacs{13.85.Hd}
\maketitle
\section{Introduction}
The Quark Gluon Plasma (QGP), an exotic state of matter of de-confined quarks and gluons, was predicted \cite {ref01} to
constitute several superdense astrophysical and cosmological situations like, the core of Neutron Stars and a few micro-second 
old Universe after the Big-bang. High energy collisions of heavy ions, forming matter of finite extension ( popularly termed as a "fireball" ) 
at extreme temperature and density, had been considered \cite {ref02, ref03} as the potential means for creating QGP in the 
laboratory. The QGP could indeed be formed in the laboratory, as has been confirmed \cite {ref04, ref05, ref06, ref07} by 
experiments with collisions of heavy ions at the Relativistic Heavy Ion Collider (RHIC) at the Brookhaven National 
Laboratory subsequent to the CERN-declaration \cite{ref08} of indications of formation of QGP- like new state of matter at
Super Proton Synchrotron (SPS). Precisely, the "fireball" created in heavy-ion collisions at RHIC is a fluid-like system of 
strongly interacting quark gluon plasma or sQGP, as has been characterized primarily by the collective flow of the produced 
final state particles. In extracting signals of QGP in heavy-ion collisions, in some cases, the data of proton-proton ($pp$) collisions 
at the same energy serve the baseline, as the $pp$ collisions are not expected to form similar hydrodynamic system according 
to the general understanding based on most of the theoretical and phenomenological models, in practice. However, there 
had always been a different school of thought \cite {ref09, ref10, ref11, ref12} that nurtured the possibility of the formation of similar system 
of smaller size in $pp$ collisions. Much earlier than the LHC - era, high multiplicity $p\bar p$ events in experiments \cite{ref13, ref14} 
at the Super Proton Synchrotron (SPS) at CERN led to the consideration of the occurrence of high energy density events and motivated 
searches for evidence of hadronic de-confinement in $p\bar p$ collisions at $\sqrt {s} $ = 0.54 TeV at SPS \cite {ref09} and at $\sqrt {s} $ 
= 1.8 TeV  \cite {ref10, ref11} at the Tevatron, Fermilab. The Analysis of transverse momentum data of the Tevatron revealed \cite {ref10} 
common radial flow velocity for meson and anti-baryon, which had been attributed to as an evidence for collectivity due to the formation of QGP. 
The $pp$ collisions at LHC energies have resulted a number of unexpected observations, having close resemblance to the signals for the 
hydrodynamic system formed in relativistic heavy-ion collisions. In this article, we address collectivity in $pp$ collisions at LHC in terms of 
transverse radial flow. 

\section{Motivation}
\label{}
One of the important findings in $pp$ - collisions at LHC, in relation to the interest of the present work, is "the ridge" 
as observed \cite {ref16} in high multiplicity events at $\sqrt {s}$= 7 TeV, 
while a similar "ridge" structure observed \cite {ref07} in the heavy-ion collisions at RHIC has been attributed to the hydrodynamical 
evolution of the system formed in the collisions. The other important observation, in this context, is related to the study of intensity 
interferometry \cite {ref17} or the Hanbury-Brown-Twiss (HBT) correlations. The dependencies of the HBT-radii of the source of emission 
of particles on multiplicity and pair transverse momentum for high multiplicity $pp$ events at LHC show  \cite {ref18, ref19} similar behavior as 
seen in the collisions of heavy ions at RHIC where the observations have been interpreted \cite {ref20} as signatures of collective 
behavior of the source. Because of these striking observations and availability of large statistics of high multiplicity, high energy density 
$pp$ events, particularly at  $\sqrt {s} $ = 7 TeV, there have been several initiatives \cite {ref21,ref22, ref23, ref24, ref25, ref25, ref26, 
ref27, ref28} either in explaining $pp$ data in the light of collective models or in predicting collective phenomena in $pp$ collisions at LHC 
energies. Collective behavior could also be studied in terms of other prescribed observables like, direct photons at low 
transverse momentum \cite {ref26}, strange baryon to meson ratio \cite {ref27} and the transverse radial flow velocity 
 \cite {ref10, ref29, ref30}. A recent analysis in terms of strange baryon to meson ratio \cite {ref31} for the $pp$ collision 
 data at $\sqrt {s}$ = 0.9 and 7 TeV by the CMS experiment, however, does not reveal the related signature of collectivity,  
 It is worth noting at this point that beside being considered as manifestation of fluid dynamical 
behavior \cite {ref24, ref25}, due to de - confinement of quarks, the appearance of the "ridge" structure has been explained \cite {ref32} also 
in the framework of the Color Glass Condensate (CGC) \cite {ref33}. 

 \begin{center}
\begin{figure}[htbp]
\includegraphics[scale=0.45]{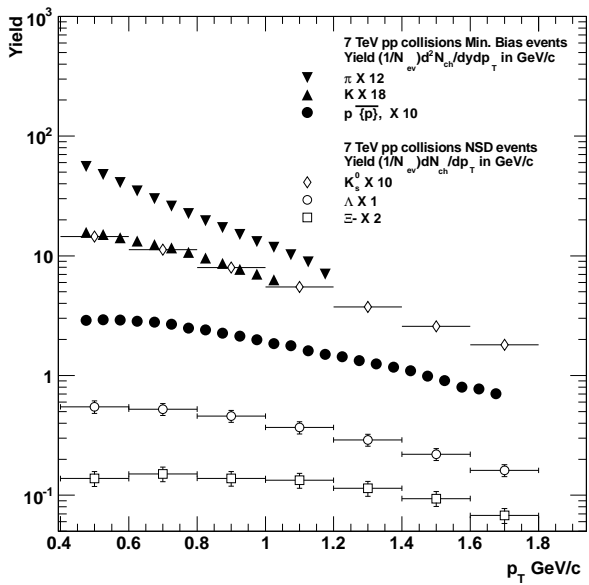}
\caption{The transverse momentum spectra for $\pi^\pm$, $K^\pm$, $p$($\bar p$) within rapidity range $|y|<1$ for minimum bias events of
$pp$ collisions and for $K_{s}^0$, $\Lambda$($\bar \Lambda$) and $\Xi^-$ within rapidity range $|y|<2$ for the non-single diffractive events 
of $pp$ collisions as measured by the CMS experiment \cite {ref34, ref31} at $\sqrt {s}$ = 7 TeV. The uncorrelated statistical and systematic 
uncertainties have been added in quadrature.}
\label{fig:pt_7000MB} 
\end{figure}
\end{center}

In the given scenario, primarily, the relative flattening of the low $p_{T}$ part of the $p_{T}$ - spectra of identified particles of 
higher masses in $pp$ collisions at LHC, as measured by the CMS experiment \cite {ref31, ref34}, motivates us to search for 
the signal of collective transverse radial flow in $pp$ collisions at LHC. It is worth mentioning, here, that in heavy-ion 
collisions at SPS, RHIC and LHC where formation of hydrodynamic system is evident through 
several signatures, the phenomenon of flattening of spectra of particles of heavy masses is attributed to the common transverse radial flow 
velocity. 

As the collectivity is inherently 
an effect associated with the "soft" particle production, the relevant signals can be better extracted from the particles populating the low - $p_{T}$ 
part of the $p_{T}$ - spectra. The low - $p_{T}$ part of the $p_{T}$ - spectra of  $\pi^\pm$, $K^\pm$,  $p$($\bar p$) from the minimum bias 
events and $K_{s}^0$, $\Lambda$, $\Xi^{-} $ from the NSD events, both at $\sqrt {s}$ = 7 TeV, are plotted in Figure~\ref{fig:pt_7000MB}. 
It may be noted that according to an estimation by the the PYTHIA event generator, in order to compare the minimum-bias measurement 
with non-single-diffractive event sample, the particle yields from the minimum-bias events should be divided \cite {ref34} by a factor 0.91 
for collisions at $\sqrt {s}$ = 7 TeV. Also, as the two measurements are in different $|y|$ - ranges, for exact comparison, one needs to 
normalize the yield appropriately. 
But, as the aim for this particular plot is to compare the slopes of the spectra, the measured \cite {ref31, ref34} spectra are multiplied with 
arbitrary factors for better illustration in the plot and the flattening of spectra with increasing mass of the particles is clearly visible.
  
\section{Methodology and Background}
\label{}
In the present work, we  search for the collective transverse radial flow in $pp$ collisions at LHC energies,
following the hydrodynamics-motivated Boltzmann - Gibbs blast - wave ( BGBW ) model \cite {ref35} which is well tested in the study of transverse radial 
flow for the established hydrodynamic systems formed in heavy-ion collisions. Application of the same formalism helps comparing the systems 
formed in different types of collisions.

\subsection{Methodology: The BGBW Model}
\label{}

The blast-wave model assumes that the particles in the system produced in the collision are locally thermalized and the system expands 
collectively with a common velocity field undergoing an instantaneous common freeze-out at a kinetic freeze-out temperature ($T_{kin}$) 
and a common transverse radial flow velocity ($\beta$) at the freeze-out surface. 

Assuming the hard-sphere particle source of uniform density, the transverse momentum spectral shape, in the BGBW model, is given by, 
\begin{equation}
\frac{dN}{p_{T}dp_{T}} \propto \int_{0}^{R} rdr \ m_{T} \ \bold I_{o} \left(\frac{p_{T}Sinh \ \rho}{T_{kin}}\right)  \bold K_{1}\left(\frac{p_{T}Cosh \ \rho}{T_{kin}}\right)
\end{equation}  
where $\rho$ = $ tan h^{-1}\beta$, $\bold I_{0}$ and $\bold K_{1}$ are modified Bessel functions. The flow velocity profile is given by,
\begin{equation}
\beta = \beta_{s}(\frac{r}{R})^{n}
\end{equation}  
where $\beta_{s}$ is the surface velocity and $r/R$ is the relative radial position in the thermal source. The average transverse flow velocity,
$\langle\beta\rangle$ is given by, $\langle\beta\rangle  =  \frac{2}{(2+n)}$$\beta_{s}$.

Though the blast-wave formalism is not a full hydrodynamic calculations, it is a useful tool for comparison of spectra-data from different 
experiments in terms of a few parameters. 
The $p_{T}$ spectra for different identified produced particles in $pp$ collisions have been 
measured at LHC at different centre-of-mass energies, $\sqrt {s}$ and for different classes of events, depending on event multiplicity.

 To compare the measured spectra with the model description by way of fitting, one needs to fix a conservative criterion to estimate the goodness 
 of the fits, or in other words, for acceptance / rejection of the model with certain level of significance. For our present study of identified particle 
 spectra in $pp$- collisions 
 in terms of BGBW model we use the Chi-square ($\chi^2$) test and compare the minimized $\chi^2$ - value with the critical value ($\chi^2_c$) 
 of the $\chi^2$ - distribution, a function of number of degrees of freedom, for the usual significance level of 0.05. If the calculated value of $\chi^2$ 
 is less than the $\chi^2_c$, there likely to exist not so significant difference between the model description of the spectra and the measured ones. 
 
 An important aspect of fitting of the $p_{T}$ - spectra is the fitting-range of $p_{T}$.  While the relevant signals of collectivity can be better 
 extracted from the particles populating the low - $p_{T}$ part of the $p_{T}$ - spectra, very low - $p_{T}$ part for some species, particularly 
 for pions, are known to have a large contribution from resonance decays and so the values of the parameters extracted from the fit 
 become sensitive to the fit range used for the pions. In previous studies of pion-spectra from heavy-ion collisions, therefore, the lower $p_{T}$ - cut 
 has been chosen to be 0.5 GeV/c. For our analysis, we put similar $p_{T}$ - cut at the lower-end of spectra for all the species. At the higher side,
 the $p_{T}$-range is limited to $p_{T} < 2$ GeV/c or less, depending on the availability of the data.

As we intend to compare results of our analysis of the $p_{T}$ spectra measured in $pp$ collisions in search of transverse 
radial flow and its dependence on centre-of-mass energy of collisions, species of produced particles 
and 'centrality', with those in $AA$  and $pA$ collisions, it would be very much pertinent to discuss here the observations 
from similar analyses in $AA$ and $pA$ collisions.

\subsection{Background: BGBW in $AA$ and $pA$ Collisions}
\label{}

At SPS, the blast-wave analysis \cite {ref36} of transverse mass spectra of pions, kaons and protons from 200 A GeV sulfur - sulfur ($SS$) and 158 
A GeV lead-lead ($PbPb$) collisions data of fixed-target experiments provided evidence of collective transverse flow from the 
heavy-ion induced central collisions. Interestingly, however, the analysis \cite {ref37} of the transverse mass spectra of strange particles 
in $PbPb$ collisions at 40 and 158 A GeV/c blast-wave fits to singly and multiply strange particles separately. In fact, experimental observations
at SPS heavy-ion collisions indicate to the scenario where multi-strange hadrons are formed and get decoupled from the system earlier, at 
large energy-density, than the common freeze-out of other hadrons \cite {ref38}. The results on the centrality dependence study \cite {ref37} 
revealed that with increasing centrality the transverse flow velocity increases and the freeze-out temperature decreases. 

For the RHIC heavy ions data, the BGBW model describes the $p_{T}$ - spectra well \cite {ref39, ref40, ref41, ref42} extracting the kinetic 
freeze-out parameters, the temperature and the radial flow velocity, by simultaneous fit to the $p_{T}$ spectra of pions, kaons and protons 
for the $p_{T}$-range up to $p_{T}$ = 1.2 GeV/c. A centrality and energy dependence study \cite {ref40} for $AuAu$ collisions reveals that the 
average transverse flow velocity ( $\langle\beta\rangle$) increases and the kinetic freeze-out temperature ($T_{kin}$) decreases with both the cantre-of-mass 
energy of collisions and the centrality. For the RHIC heavy-ion collisions, like in the case of SPS heavy-ion collisions, the blast-wave fit to
the spectra of $\pi^\pm$, $K^\pm$, $p$($\bar p$) fails to reproduce \cite {ref43} spectra for $K_{s}^0$, $\Lambda$($\bar \Lambda$), $\Xi$($\bar \Xi$) and 
$\Omega$($\bar \Omega$), strengthening the viewpoint \cite {ref38, ref44} that multi-strange baryons freeze-out earlier than the other particles, possibly
due to their smaller interaction cross-section with the medium produced at the collision zone. 

At LHC, the combined fit \cite {ref45} of blast-wave to the identified particle spectra of lead-lead ($PbPb$) collisions by ALICE at $\sqrt {s_{NN}} $ 
= 2.76 TeV includes $p_{T}$-ranges, 0.5 - 1.0 GeV/c, 0.2-1.5 GeV/c and 0.3 - 3.0 GeV/c for pions, kaons and protons respectively. A centrality 
dependent study \cite {ref46} for the ALICE data shows similar behavior of  the average transverse flow velocity ( $\langle\beta\rangle$) and the kinetic 
freeze-out temperature ($T_{kin}$), as has been observed \cite {ref40} for the RHIC heavy-ion data.  

The most recent and striking observation in respect of blast-wave formalism, however, is the indication of transverse radial flow and hence the
collectivity in the $pPb$ collisions at $\sqrt {s_{NN}} $ = 5.02 TeV as has been shown \cite {ref47} by the ALICE collaboration. A simultaneous
blast-wave fit to the $p_{T}$-spectra for the $p_{T}$-ranges, 0.5 - 1.0 GeV/c, 0.2-1.5 GeV/c, 0.0-1.5 GeV/c, 0.3 - 3.0 GeV/c and 0.6-3.0 GeV/c for 
$\pi^\pm$, $K^\pm$, $K_{s}^0$, $p$($\bar p$) and $\Lambda$($\bar \Lambda$) respectively, results a strong transverse
flow velocity. Also, the dependence of the blast-wave parameters on multiplicity of different event classes for $pPb$ data is 
similar to that of centrality dependence in heavy-ion data at SPS, RHIC and LHC. The finding of stronger radial flow velocity for $pA$ 
collisions as compared to that in $AA$ collisions corroborates a very recent theoretical calculation \cite {ref48} that addresses
the question of applicability of hydrodynamics in high-multiplicity $pp$ and $pA$ collisions. According to the theory, with the
available high-multiplicity events, the hydrodynamics apparently starts to work for $pA$ and $pp$ collisions. The calculation 
\cite {ref48} predicts even stronger radial flow velocity for the high multiplicity $pp$ collisions. It is important to note that, like in 
high-multiplicity $pp$ collisions at $\sqrt {s}$ = 7 TeV, the $pPb$ at $\sqrt {s_{NN}} $ = 5.02 TeV also has revealed the `ridge' 
structure in the two-particle correlations \cite {ref49,ref49,ref50}. Another feature of spectra of identified particles which is 
considered as a signature of radial flow in $AA$ collisions has been observed \cite {ref51} by the CMS experiment in 
high-multiplicity $pPb$ collisions. The spectra of identified particles were fitted with a functional form proportional to 
$p_{T} \exp (-m_{T^{'}} / T^{'})$, where $T^{'}$ is called the inverse slope parameter. A linear increase of $T^{'}$ with mass 
of the measured identified particles, recognized as a signature of radial flow, has been observed in high-multiplicity event-classes 
of $pPb$ collisions. 

\section{Results and Discussions: BGBW in $pp$ Collisions at LHC}
\label{}
Experiments at LHC have measured $p_{T}$-spectra for different identified particles with detector setups of different geometrical 
acceptance for detecting several kinds of particles in different kinematic ranges. The scope of this phenomenological work is 
limited by the availability of published data, relevant to the objective of the work. The CMS experiment has published  $p_{T}$-spectra of two classes of 
particles, one containing  \cite {ref34, ref52} the commonly measured particles, $\pi^\pm$, $K^\pm$, $p$($\bar p$) over the rapidity, 
($y = (1/2) ln\frac{E + p_{L}}{E - p_{L}}$) range $|y|<1$ for the $pp$ collisions at $\sqrt {s}$ = 900, 2760 and 7000 GeV, while the other
 class contains  \cite {ref31, ref53} only the strange particles, strange meson, $K_{s}^0$, single-strange baryon, $\Lambda$ and the double-strange 
 baryon, $\Xi^{-} $  over the rapidity, range $|y|<2$ for the Non-single diffractive (NSD) events of $pp$ collisions at $\sqrt {s}$ = 900 and 
 7000 GeV. The $p_{T}$-spectra for  $\pi^\pm$, $K^\pm$, $p$($\bar p$) have also been measured for different classes of events having 
 selected non-overlapping range of event-multiplicity, reflecting "centrality" in $pp$ collisions. The CMS data \cite {ref52, ref53} thus 
 facilitate the centre-of-mass energy, species and centrality dependent blast-wave analysis of the $p_{T}$-spectra of identified particles 
 produced in $pp$ collisions at LHC and to compare the results with similar studies in $AA$ and $pA$ collisions. 
 
 The data of CMS experiment, that are used for this phenomenological study, have been extensively compared \cite {ref31, ref34} 
with several versions or tunes of the PYTHIA event generator. It will be pertinent to note that none of the PYTHIA tunes used in the study of the 
inclusive production of  
$\pi^\pm$, $K^\pm$ and $p$($\bar p$) in $pp$ collisions at  $\sqrt {s}$ = 0.9, 2.76 and 7 TeV \cite {ref34} provides acceptable overall description 
of data at the finer details like, the multiplicity dependence of average transverse momentum, $\langle p_{T} \rangle$ or the centre-of-mass energy, 
$\sqrt {s}$, dependence of $dN/dy$, $\langle p_{T} \rangle$ and the particle yield ratios. Similarly, in the study \cite {ref31} of $K_{s}^0$, 
$\Lambda$($\bar \Lambda$) and $\Xi^-$ also the PYTHIA tunes fails to match the increase in production of strange particles with increasing 
$\sqrt {s}$, as measured by the experiment. The discrepancy between the PYTHIA calculation and the data is more for $\Lambda$($\bar \Lambda$) 
and $\Xi^-$. 
 
Following the BGBW model \cite {ref35} as described in Section - A, we attempt to fit the blast-wave function to the $p_{T}$ - spectra for different 
sets of data, as measured \cite {ref31, ref34} by CMS experiment, keeping the kinetic freeze-out temperature ($T_{kin}$), 
the radial flow velocity ($\beta_{s}$) and the exponent ($n$) of the flow velocity profile free to produce the best possible simultaneous or 
combined fits to the data, in terms of $\chi^2$/ndf, using the MINUTE program in the ROOT analysis framework \cite {ref54}.

\subsection{BGBW fit to spectra of  $\pi^\pm$, $K^\pm$ and $p$($\bar p$) }
\label{}

We try to fit the spectra of $\pi^\pm$, $K^\pm$ and $p$($\bar p$) with the BGBW model for the minimum bias events of $pp$ collisions 
at $\sqrt {s}$ = 0.9, 2.76 and 7 TeV \cite {ref34}. The minimum bias data could not be explained satisfactorily with the blast-wave model. 

\begin{center}
\begin{figure}[htbp]
\includegraphics[scale=0.45]{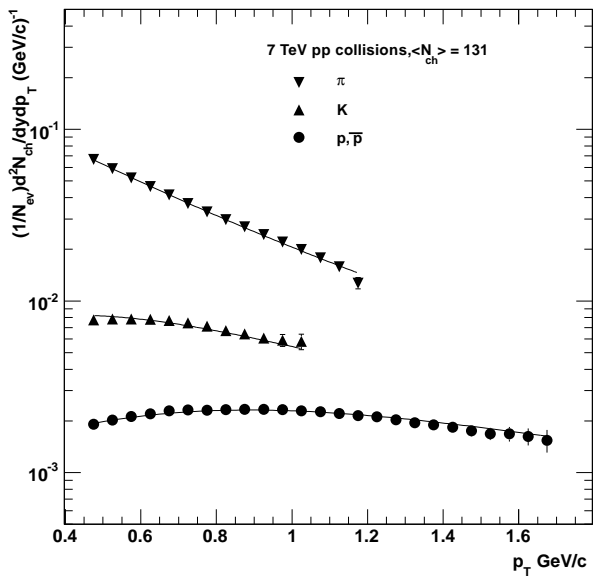}
\caption{The transverse momentum spectra for $\pi^\pm$, $K^\pm$, $p$($\bar p$) as measured by the CMS experiment \cite {ref34, ref52} at LHC 
for the event-class of average multiplicity = 131 in $pp$-collisions at $\sqrt {s}$ = 7 TeV, along with BG-blast-wave fits (solid lines). The uncorrelated 
statistical and systematic uncertainties have been added in quadrature.}
\label{fig:pt_7000m131} 
\end{figure}
\end{center}

\begin{center}
\begin{figure}[htbp]
\includegraphics[scale=0.45]{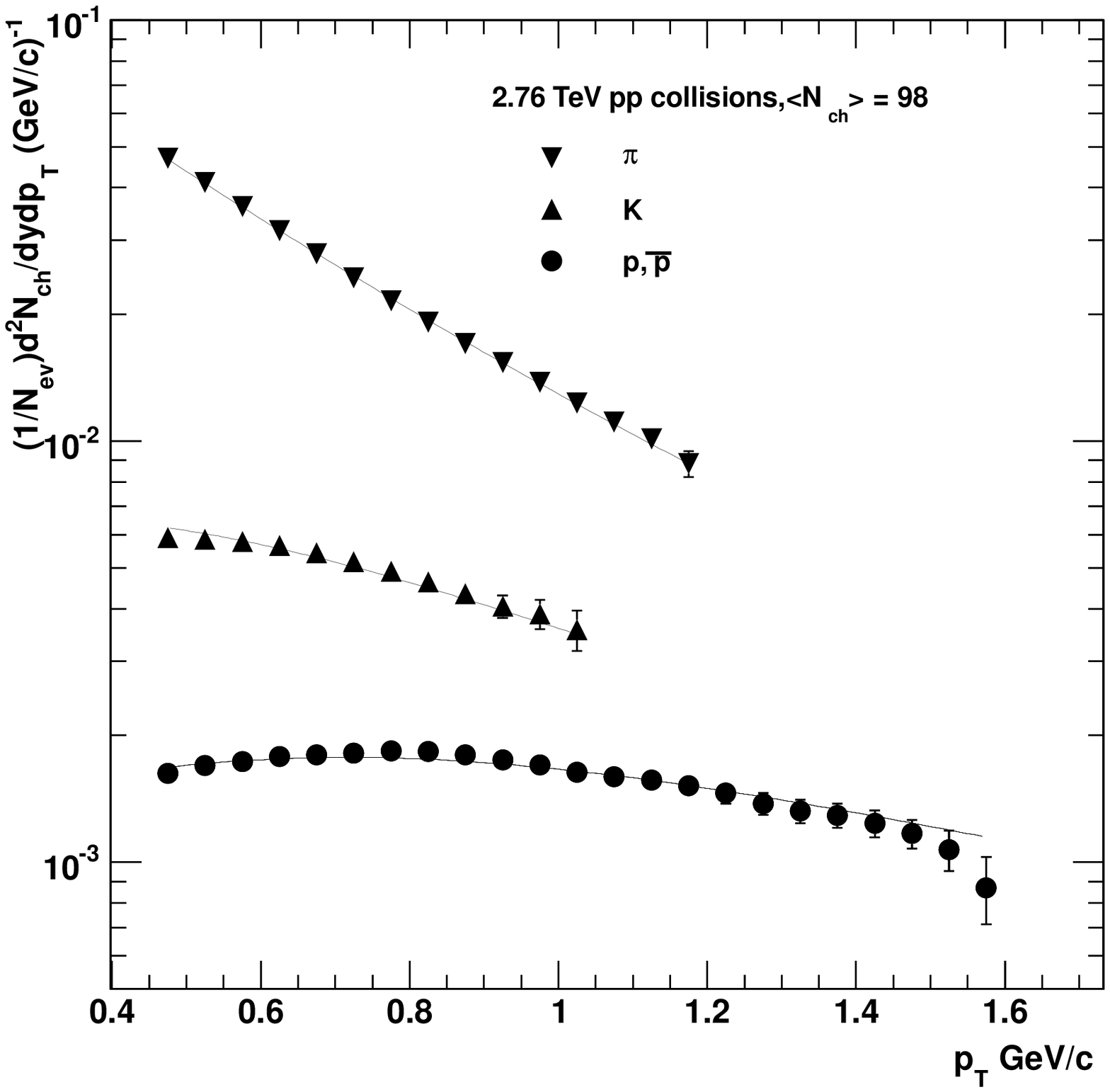}
\caption{The transverse momentum spectra for $\pi^\pm$, $K^\pm$, $p$($\bar p$) as measured by the CMS experiment \cite {ref34, ref52} at LHC 
for the event-class of average multiplicity = 98 in $pp$-collisions at $\sqrt {s}$ = 2.76TeV, along with BG-blast-wave fits (solid lines). The uncorrelated 
statistical and systematic uncertainties have been added in quadrature.}
\label{fig:pt_2760m98} 
\end{figure}
\end{center}

\begin{center}
\begin{figure}[htbp]
\includegraphics[scale=0.45]{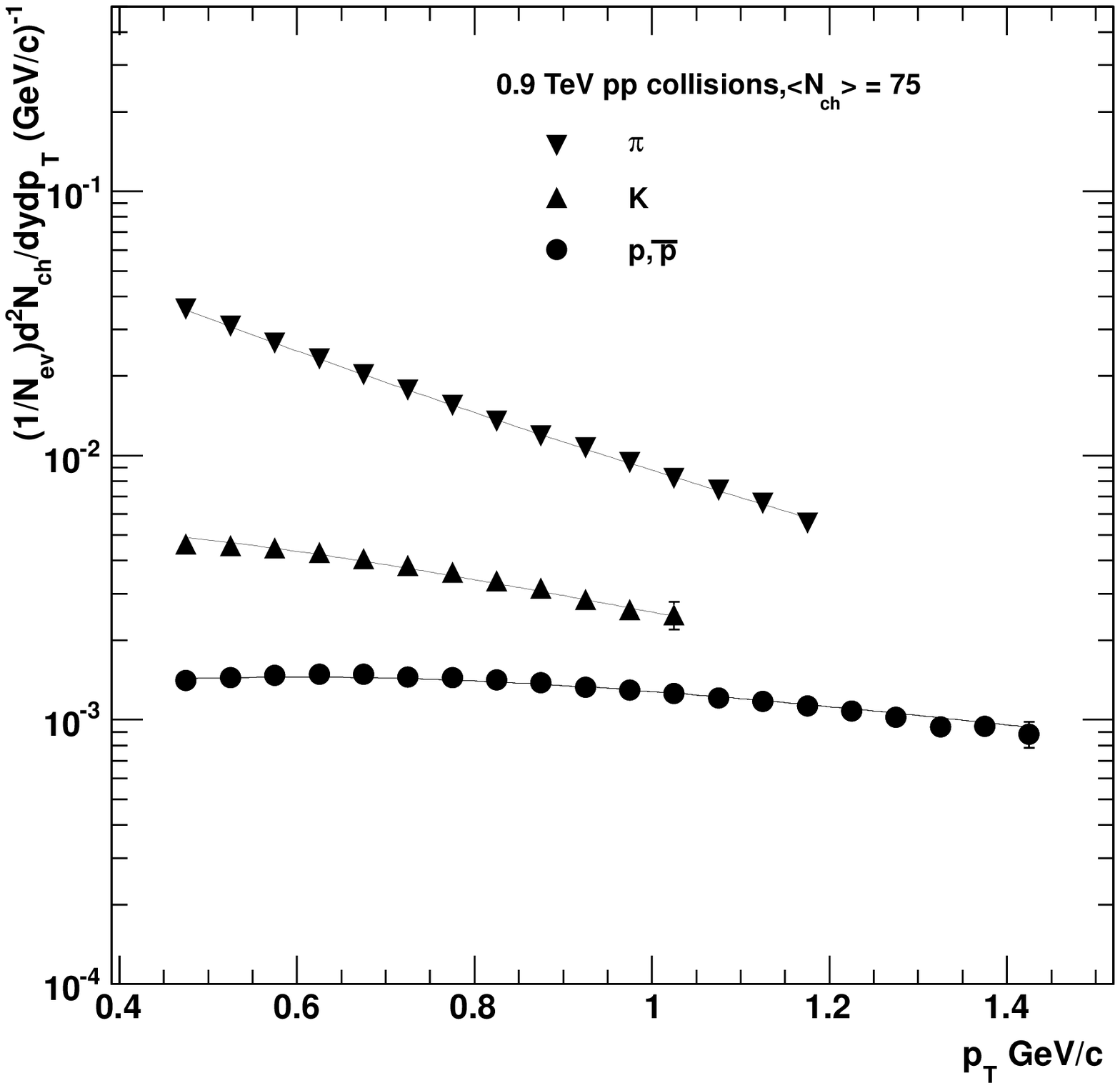}
\caption{The transverse momentum spectra for $\pi^\pm$, $K^\pm$, $p$($\bar p$) as measured by the CMS experiment \cite {ref34, ref52} at LHC for the event-class of average multiplicity = 75 in $pp$-collisions at $\sqrt {s}$ = 0.9 TeV, along with BG-blast-wave fits (solid lines). The 
uncorrelated statistical and systematic uncertainties have been added in quadrature.}
\label{fig:pt_900m75} 
\end{figure}
\end{center}

Taking advantage of the availability \cite {ref34} of the 'centrality' - dependent $p_{T}$ - spectra for $\pi^\pm$, $K^\pm$ and $p$($\bar p$) in $pp$ 
collisions and considering the fact that, so far, high multiplicity $pp$ events only have exhibited significant signatures, which could be attributed 
to the collectivity in $pp$ collisions, we continue the study and observe good matching of the BGBW model with the $p_{T}$ - spectra of events 
with high multiplicity, corresponding to very central $pp$ collisions. Figure~\ref{fig:pt_7000m131},  Figure~\ref{fig:pt_2760m98} and 
Figure~\ref{fig:pt_900m75} represent the $p_{T}$ - spectra for $\pi^\pm$, $K^\pm$ and $p$($\bar p$) as measured by CMS \cite {ref34} from 
topmost 'central' class of events, along with simultaneous BGBW fits, for the $pp$ - collisions at $\sqrt {s}$ = 7, 2.76 and 0.9 TeV, respectively. 
Of course, for several other classes of events of high-multiplicity, particularly for $\sqrt {s}$ = 7 TeV, similar spectra can also be well described 
by the BGBW model.

 A sensitive input to the blast-wave model is the transverse radius ($R$) of the source of particle emission at freeze-out. Ideally, the values 
 of $R$ could be obtained from the pair transverse momentum dependent HBT-radius. The HBT-radius corresponding to the lowest
 value of pair transverse momentum may be considered as the radius of the source of the emission close to the freeze-out. The 
 ALICE \cite {ref18} and the CMS experiments \cite {ref19} have shown that, like in the case of heavy-ion collisions, the radius of 
 the source of emission of particle in $pp$ collisions at LHC also shows the pair transverse momentum and the multiplicity dependence. 
 As the selection of multiplicity classes used in this study is not the same as that in Ref. \cite {ref18} or \cite {ref19} and as the published
 data of pair transverse momentum and multiplicity dependent radius (for event classes of low multiplicities) do not produce any
 scaling behavior with multiplicity, we calculate $R$ for different event classes with different $\langle N_{ch} \rangle$ in $|\eta| <2.4$, as used 
 in this study, from the relation, $R(\langle N_{ch} \rangle) = a. \langle N_{ch} \rangle^{1/3}$ where $a = 0.597 \pm 0.009 (stat.) \pm 0.057(syst.)$ 
 fm at 0.9 TeV and $a = 0.612 \pm 0.007 (stat.) \pm 0.068(syst.)$ fm at 7 TeV, as have been parameterized \cite {ref19} by the CMS experiment from
 the measurement of radius of source of emission as a function of average charged particle multiplicity in the range $|\eta| <2.4$. 
 For the 2.76 TeV $pp$ collisions, we estimate the radius of the source of emission by interpolation of radii at 0.9 and 7 TeV for 
 respective $\langle N_{ch} \rangle$.

The kinetic freeze-out temperature ($T_{kin}$), the average radial flow velocity  ($\langle\beta\rangle$) at the freeze-out surface, 
and the exponent ($n$) as obtained by simultaneous fit by BGBW for different classes of events, indicating different centrality, and for 
different centre-of-mass energies ($\sqrt {s}$), along with respective $\chi^2/ n.d.f$ are tabulated in Table - I. Obviously, the table
contains the values of parameters for those event classes which pass the set criterion of good-fit.
 
We have noted that for $AA$ and $pA$ collisions, at a given centre-of-mass energy, $\langle \beta \rangle$ increases and $T_{kin}$ decreases 
with increasing average event multiplicity or the 'centrality'. The values of $\langle \beta \rangle$, as tabulated, from our analysis of $pp$ collisions 
data show similar dependency for $pp$ events in $\sqrt {s}$ = 2.76 and 7 TeV. For the most 'central' or the topmost class of high-multiplicity $pp$ 
events, the average radial flow velocity, ($<\beta>$) increases and the kinetic freeze-out temperature, ($T_{kin}$) remains almost the same with 
the increase in the centre-of-mass energy of collisions.

\begin{table}[h]
\begin{center}
\label{tab}
\scalebox{0.8}{
\begin{tabular}{|c|c|c|c|c|c|}
\hline
$\sqrt {s} (TeV)$&$\langle N_{ch} \rangle$& $T_{kin} (MeV) $& $\langle \beta \rangle$& $n$&$\chi^2/n.d.f$ \\
\hline
$ 0.9$ &$75$&$106.43\pm 0.10$ & $ 0.745\pm 0.004$ & $ 0.584\pm 0.010$&$0.29$\\
\hline
$ 2.76$ &$86$&$115.55\pm 0.11$ & $ 0.742\pm 0.005$ & $ 0.605\pm 0.007$&$1.25$\\
\hline
$ 2.76$ &$98$&$110.39\pm 0.13$ & $ 0.769\pm 0.005$ & $ 0.521\pm 0.009$&$0.43$\\
\hline
$ 7$ &$98$&$115.57\pm 0.11$ & $ 0.766\pm 0.004$ & $ 0.540\pm 0.006$&$1.02$\\
\hline
$ 7$ &$109$&$113.09\pm 0.12$ & $ 0.779\pm 0.004$ & $ 0.503\pm 0.006$&$0.61$\\
\hline
$ 7$ &$120$&$110.84\pm 0.15$ & $ 0.790\pm 0.004$ & $ 0.480\pm 0.006$&$0.34$\\
\hline
$ 7$ &$131$&$104.29\pm 0.15$ & $ 0.809\pm 0.005$ & $ 0.436\pm 0.005$&$0.44$\\
\hline
\end{tabular}
}
\caption{$T_{kin}$, $\langle\beta\rangle$ and $n$, the parameters of the the BGBW, obtained from the simultaneous fit to the published 
\cite {ref52} spectra of  $\pi^\pm$, $K^\pm$ and $p$($\bar p$) and respective $\chi^2 /n.d.f$ for $pp$ collisions at $\sqrt {s}$ = 0.9, 2.76 
and 7 TeV for different event classes depending on average multiplicity, $\langle N_{ch} \rangle$, in the range $|\eta| < 2.4$.}
\end{center}
\end{table}

\begin{center}
\begin{figure}[htbp]
\includegraphics[scale=0.45]{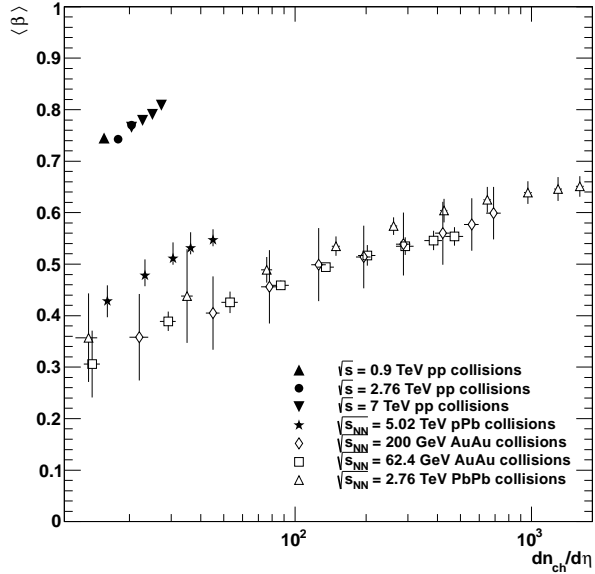}
\caption{The $\sqrt {s}$ and centrality ($dn_{ch}/d\eta$) dependence of transverse radial velocity, $\langle\beta\rangle$, as obtained by simultaneous fits in the BGBW framework to the published \cite {ref34, ref52} spectra of  $\pi^\pm$, $K^\pm$ and $p$($\bar p$) in $pp$ 
collisions at LHC is compared with results from similar analysis for $AuAu$ collisions at RHIC \cite {ref40}, $PbPb$ and $pPb$ collisions 
at LHC \cite {ref46, ref47}. }
\label{fig:betaVSdndy} 
\end{figure}
\end{center}

\begin{center}
\begin{figure}[htbp]
\includegraphics[scale=0.45]{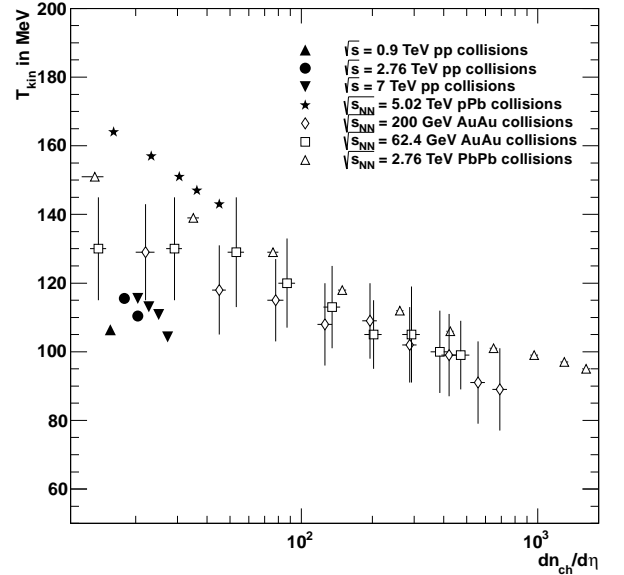}
\caption{The $\sqrt {s}$ and centrality ($dn_{ch}/d\eta$) dependence of kinetic freeze-out temperature, $T_{kin}$, as obtained by simultaneous fits in the 
BGBW framework to the published \cite {ref34, ref52} spectra of  $\pi^\pm$, $K^\pm$ and $p$($\bar p$) in $pp$ collisions at LHC is compared with results
from similar analysis for $AuAu$ collisions at RHIC \cite {ref40}, $PbPb$ and $pPb$ collisions at LHC \cite {ref46, ref47}. }
\label{fig:tkinVSdndy} 
\end{figure}
\end{center}

The comparison of the $\sqrt {s}$ and 'centrality' dependences of the BW-parameters for the $pp$ collisions with those for $pA$ and $AA$ 
collisions can be better visualized in Fig.~\ref{fig:betaVSdndy} and Fig.~\ref{fig:tkinVSdndy}.

In a hydrodynamic picture, the collective radial flow is generated due to the pressure gradient in the system. The larger the pressure gradient, the
greater is the radial flow velocity. The profile of the pressure gradient is determined by the initial energy density and 
the spatial size or lifetime of the system formed in the collision. A higher initial energy density and smaller size of the system generates larger pressure gradient. 
The high-multiplicity $pp$ collisions at LHC energies thus likely to generate larger pressure gradient as compared to the heavy-ion collisions 
at SPS, RHIC and LHC or the $pPb$ collisions at LHC at similar centrality ($dn_{ch}/d\eta$). Also, the pressure gradient in $PbPb$ collisions 
at  $\sqrt {s_{NN}} $ = 2.76 TeV is larger than that in $AuAu$ collisions at RHIC energies and similarly, that is larger for $pPb$ collisions at  
$\sqrt {s_{NN}} $ = 5.02 TeV as compared to $PbPb$  collisions at $\sqrt {s_{NN}} $ = 2.76 TeV. Thus the  $\sqrt {s}$, the centrality ($dn_{ch}/d\eta$) 
and the system ( $pp$, $pA$ and $AA$) dependences of transverse radial velocity, $\langle\beta\rangle$, as a consequence of stronger radial 
gradients  \cite {ref48}, as shown in Fig.~\ref{fig:betaVSdndy} appear consistent. 

The Fig.~\ref{fig:tkinVSdndy} shows that for each of the systems $pp$, $pA$ and $AA$, the kinetic freeze-out temperature, $T_{kin}$, decreases
with increasing centrality.  However, while the $pPb$ system freezes out at larger $T_{kin}$ compared to the $AuAu$ or the $PbPb$ systems, 
the $T_{kin}$ in $pp$ system is lower than both the $pA$ and $AA$ systems at comparable $dn_{ch}/d\eta$. The $T_{kin}$ for the high-multiplicity 
$pp$ collisions is rather comparable with that for the central $AA$ collisions. Theoretical calculations in Ref. \cite {ref48} show 
that, at a given $T_{kin}$, even though the absolute sizes and multiplicities in central $AA$ collisions are quite different from the type of 
multiplicity-class of $pp$ collisions studied here, the $pp$ system gets more `explosive' than the $AA$ system and so, according to the hydrodynamic
picture, the transverse collective flow velocity, a function of $t$ (the time from the initial timeline) and  $r$ (radial distance from the centre of the fireball),  
for the smaller system reaches a larger value at the freeze-out surface. 

\subsection{BGBW fit to spectra of  $K_{s}^0$, $\Lambda$($\bar \Lambda$) and $\Xi^-$}
\label{}
\begin{center}
\begin{figure}[htbp]
\includegraphics[scale=0.45]{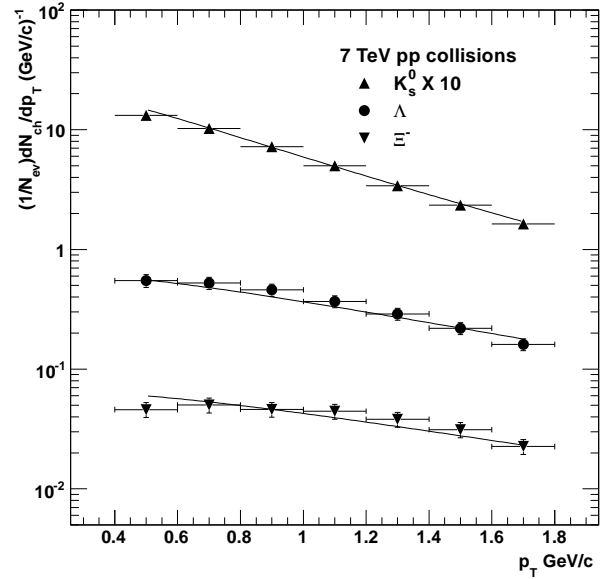}
\caption{The transverse momentum spectra for the strange meson, $K_{s}^0$, single-strange baryon, $\Lambda$ and the double-strange 
baryon, $\Xi^-$), as measured by the CMS experiment \cite {ref31, ref53} at LHC in the minimum bias $pp$-collisions at $\sqrt {s}$ = 7 TeV, along 
with BG-blast-wave fits (solid lines). The uncorrelated statistical and systematic uncertainties have been added in quadrature.}
\label{fig:pt_7000} 
\end{figure}
\end{center}

\begin{center}
\begin{figure}[htbp]
\includegraphics[scale=0.45]{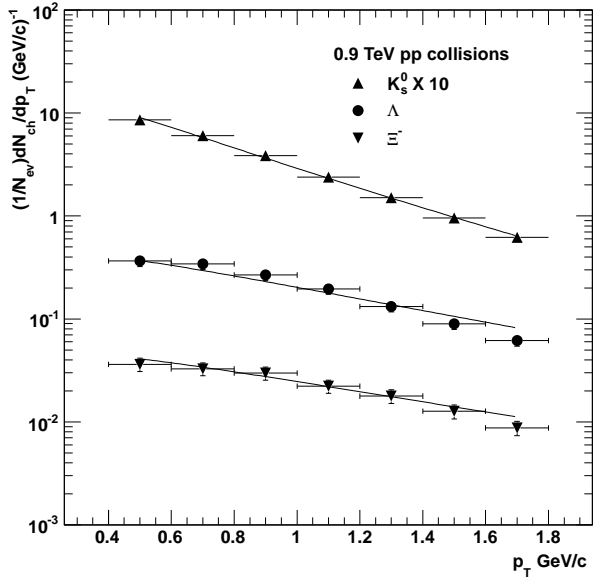}
\caption{The transverse momentum spectra for the strange meson, $K_{s}^0$, single-strange baryon, $\Lambda$ and the double-strange 
baryon, $\Xi^-$), as measured by the CMS experiment \cite {ref31, ref53} at LHC in the minimum bias $pp$-collisions at $\sqrt {s}$ = 0.9 TeV, along 
with BG-blast-wave fits (solid lines). The uncorrelated statistical and systematic uncertainties have been added in quadrature.}
\label{fig:pt_900} 
\end{figure}
\end{center}

Although the spectra of $\pi^\pm$, $K^\pm$ and $p$($\bar p$) from minimum bias $pp$ collisions could not be described by the BGBW 
model, we attempt to apply the blast-wave model in describing exclusive spectra of strange particles as measured \cite {ref31, ref53} by the 
CMS experiment from the NSD events in the rapidity-range $|y|<2$. The strange particle spectra, parts of which are drawn in 
Fig.~\ref{fig:pt_7000} and Fig.~\ref{fig:pt_900}, have been obtained by reconstructing their decays: strange meson, 
$K_{s}^0 \rightarrow \pi^{+}\pi^{-}$, single-strange baryon, $\Lambda \rightarrow p\pi^{-}$ and the double-strange 
baryon, $\Xi^{-} \rightarrow \Lambda \pi^{-}$. 

\begin{table}[h]
\begin{center}
\label{tab}
\begin{tabular}{|c|c|c|c|c|}
\hline
$\sqrt {s} (TeV)$& $T_{kin} (MeV) $& $\langle\beta\rangle$& $n$&$\chi^2/n.d.f$ \\
\hline
$ 7$ &$149\pm 0.59$ & $ 0.62\pm 0.006$ & $ 1.0\pm 0.02$&$0.85$\\
\hline
$ 0.9$ &$140\pm 0.53$ & $ 0.54\pm 0.01$ &$1.27\pm0.12$& $0.62$ \\
\hline
\end{tabular}
\caption{$T_{kin}$, $\langle \beta\rangle$ and $n$, the parameters of the the BGBW, obtained from the simultaneous fit to the published 
\cite {ref31, ref53} spectra of the strange meson, $K_{s}^0$, single-strange baryon, $\Lambda$ and the double-strange 
baryon, $\Xi^-$ and the respective $\chi^2 /n.d.f$ for $pp$ collisions at $\sqrt {s}$ = 0.9 and 7 TeV. }
\end{center}
\end{table}

The BGBW function fits to the data of all strange particles well in the $p_{T}$ - range, 0.5 $>p_{T} <$ 1.8 GeV/c as can be seen in 
Fig.~\ref{fig:pt_7000} and Fig.~\ref{fig:pt_900}. The kinetic freeze-out temperature ($T_{kin}$), the average radial flow velocity  
($\langle\beta\rangle$) at the freeze-out surface, and the exponent ($n$) as obtained by fitting the identified particle spectra, along with respective 
$\chi^2/ n.d.f$ are tabulated in Table - II. The statistical and systematic uncertainties, as quoted with the spectra data \cite {ref53}, have been 
added in quadrature.

The species dependence of blast-wave study for $pp$ collisions could not be studied on the same footing, as the published spectra 
of $\pi^\pm$, $K^\pm$ and $p$($\bar p$) and those for all strange particles, $K_{s}^0$, $\Lambda$ and $\Xi^-$ were measured in
different $|y|$ range and for different classes of events. Further, the spectra for $K_{s}^0$, $\Lambda$ and $\Xi^-$ for multiplicity dependent 
event classes are not available. Nevertheless, different response of spectra of the two groups of particles to the BGBW description, 
as has been observed in heavy-ion collisions \cite {ref38, ref43}, continues to be seen in $pp$ collisions also. 

In $pp$ collisions, while the measured spectra of 
$\pi^\pm$, $K^\pm$ and $p$($\bar p$) from the minimum bias events of $pp$ collisions at LHC at $\sqrt {s}$ = 0.9, 2.76 and 7 TeV do not fit the 
BGBW, the spectra for $K_{s}^0$, $\Lambda$ and $\Xi^-$ form the NSD events at $\sqrt {s}$ = 0.9 and 7 TeV could be described by the blast-wave 
model. The observation of good description of spectra for $K_{s}^0$, $\Lambda$ and $\Xi^-$ by BGBW could be naively linked to the viewpoint of 
the early freeze-out of multi-strange particles, as has been interpreted from the SPS and RHIC heavy-ion data analysis. 

\section{Summary and Remarks}
\label{}

We search for the transverse radial flow in $pp$ collisions by analyzing identified particle spectra in the framework of Boltzmann-Gibbs blast-wave, 
a known method of characterizing collectivity in relativistic collisions of heavy nuclei.

\textbullet {The BGBW model cannot describe the measured spectra of $\pi^\pm$, $K^\pm$ and $p$($\bar p$) from the minimum bias events of $pp$ 
collisions at LHC at $\sqrt {s}$ = 0.9, 2.76 and 7 TeV. }

\textbullet {Reasonably good simultaneous fits of the BGBW description to the transverse momentum spectra 
of $\pi^\pm$, $K^\pm$ and $p$($\bar p$) in high multiplicity events in $pp$ collisions at $\sqrt {s}$ = 0.9, 2.76 and 7 TeV at LHC indicate 
to the formation of collective medium in high-multiplicity events in $pp$ collisions at LHC.} 

\textbullet {The dependency of the average transverse radial flow velocity on mean event multiplicity or the 'centrality' for high-multiplicity $pp$ events at 
$\sqrt {s}$ = 7 TeV is strikingly similar to that for the $AA$ collisions, where collectivity due to de-confinement of quarks is reasonably 
established.} 

\textbullet {The stronger radial flow velocity for high-multiplicity $pp$ collisions as compared to $pA$ and $AA$ collisions is consistent 
with the hydrodynamic picture.}    

\textbullet {Different responses of the spectra of the two groups of particles, 1) $\pi^\pm$, $K^\pm$, $p$($\bar p$) and 2) $K_{s}^0$, 
$\Lambda$, $\Xi^-$ to the BGBW model has been observed in the $AA$ collisions at SPS and RHIC also.} 

The findings are encouraging for continuing the search for transverse radial flow or the collectivity, in general, 
in $pp$ collisions in full hydrodynamic calculations. 

\section{Acknowledgement}
PG acknowledges useful discussions with Asis Chaudhuri, VECC, Kolkata.


\begin{thebibliography}{99}
\bibitem{ref01} J.C. Collins and M.J. Perry, Phys. Rev. Lett. {\bf 34},1353 (1975).
\bibitem{ref02} E.V. Shuryak, Phys. Repts. {\bf 61}, 71(1980).
\bibitem{ref03} M.J. Tannenbaum, Rep. Prog. Phys. {\bf 69}, 2005 (2006).
\bibitem{ref04} I. Arsene et al., BRAHMS Collaboration, Nucl. Phys. {\bf A757}, 1 (2005).
\bibitem{ref05} K. Adcox et al., PHENIX Collaboration, Nucl. Phys. {\bf A757}, 184 (2005).
\bibitem{ref06} B.B. Back et al., PHOBOS Collaboration, Nucl. Phys.  {\bf A757}, 28 (2005).
\bibitem{ref07} J. Adams et al., STAR Collaboration, Nucl. Phys. {\bf A757}, 102 (2005).
\bibitem{ref08} CERN press release, http://pressold.web.cern.ch/PressOld/
Releases00/PR01.00EQuarkGluonMatter.html, 2000.
\bibitem{ref09} L. Van Hove Phys. Lett.{\bf B118}, 138 (1982).
\bibitem{ref10} P. Levai and B. Muller Phys. Rev. Lett. {\bf 67}, 1519 (1991).
\bibitem{ref11} T. Alexopoulos et al., Phys. Lett.{\bf B528}, 43 (2002).
\bibitem{ref12} R.M. Weiner Int.J.Mod.Phys {\bf E15}, 37 (2006).
\bibitem{ref13} K. Alpgard et al.,  UA5 Collaboration, Phys. Lett. {\bf B107}, 310 (1981).
\bibitem{ref14} G. Arnison et al., UA1 Collaboration), Phys. Lett. {\bf B107}, 320 (1981).
\bibitem{ref15} V. Khachatryan et al., CMS Collaboration, J. High Energy Phys. {bf 09 }, 091 (2010).
\bibitem{ref16} J. Adams et al., STAR Collaboration, Phys. Rev. Lett. {\bf 95}, 152301 (2005).
\bibitem{ref17} G. Goldhaber et al., Phys. Rev. {\bf 120}, 300 (1960).
\bibitem{ref18} K. Aamodt et al., ALICE Collaboration, Phys. Rev.{\bf D84}, 112004 (2011).
\bibitem{ref19} V. Khachatryan et al., CMS Collaboration, J. High Energy Phys. {\bf 05}, 029 (2011). 
\bibitem{ref20} J. Adams et al., STAR Collaboration, Phys. Rev.{\bf C71}, 044906 (2005).
\bibitem{ref21} S.K. Prasad et al., Phys. Rev.{\bf C82}, 024909 (2010).
\bibitem{ref22} J. Casalderrey-Solana and U.A. Wiedemann, Phys. Rev. Lett. {\bf 104}, 102301 (2010).
\bibitem{ref23} D. d'Enterria  et al., Eur. Phys. J.{\bf C66}, 173 (2010).
\bibitem{ref24} P. Bozek, Eur. Phys. J.{\bf C71}, 1530 (2011).
\bibitem{ref25} K. Werner, L. Karpenko and T. Pierog, Phys. Rev. Lett. {\bf 106}, 122004 (2011).
\bibitem{ref26} F. Liu and K. Werner, Phys. Rev. Lett. {\bf 106}, 242301 (2011).
\bibitem{ref27} V. Topor et al., Phys. Rev. {\bf C86}, 044902 (2012).   
\bibitem{ref28} K. Werner et al., Phys. Rev. {\bf C82}, 044904 (2010).
\bibitem{ref29} K. Werner et al., Phys. Rev.{\bf C83}, 044915 (2011).  
\bibitem{ref30} M. Floris et al., ALICE Collaboration, J.Phys. G {\bf G270}, Conf. Series 012046 (2011).
\bibitem{ref31} V. Khachatryan et al., CMS Collaboration, J. High Energy Phys. {\bf 05 }, 064 (2011). 
\bibitem{ref32} A. Dumitru et al., Phys. Lett.{\bf B697}, 21(2011).
\bibitem{ref33} L. McLerran arXiv:hep-ph/0104285v2 (2001).
\bibitem{ref34} V. Khachatryan et al., CMS Collaboration, Euro. Phys. J. {\bf C}, 72:2164 (2012). 
\bibitem{ref35} E. Schnedermann, J. Sollfrank and U.W. Heinz, Phys. Rev.{\bf C48}, 2462 (1993). 
\bibitem{ref36} I.G. Bearden et al., NA44 Collaboration, Phys. Rev. Lett. {\bf 78}, 2080 (1997). 
\bibitem{ref37} G.E. Bruno et al., NA49 Collaboration, J. Phys. {\bf G31},S127 (2005).
\bibitem{ref38} H. Van Hecke, H.Sorge and N. Xu,  Phys. Rev. Lett. {\bf 81}, 5764 (1998). 
\bibitem{ref39} J. Adaams et al., STAR Collaboration, Phys. Rev. Lett. {\bf 92}, 112301 (2004).
\bibitem{ref40} B.I. Abelev et al., STAR Collaboration, Phys. Rev. {\bf C79}, 034909 (2009).
\bibitem{ref41} F. Retiere and M.A. Lisa, Phys. Rev.{\bf C70}, 044907 (2004).
\bibitem{ref42} S. Choi and K.S. Lee, Phys. Rev.{\bf C84}, 064905 (2011). 
\bibitem{ref43} O. Barannikova for STAR Collaboration arXiv:nucl-ex/0408022v1(2004).
\bibitem{ref44} J. Adams et al., STAR Collaboration, Phys. Rev. Lett. {\bf 92}, 182301 (2004).
\bibitem{ref45} K. Aamodt  et al., ALICE Collaboration, Phys. Rev. Lett. {\bf 109}, 252301 (2012).
\bibitem{ref46} K. Aamodt  et al., ALICE Collaboration, Phys. Rev.{\bf C88}, 044910 (2013).
\bibitem{ref47} K. Aamodt  et al., ALICE Collaboration, Phys. Lett. {\bf B719}, 29 (2013).   
\bibitem{ref48} E. Shuryak and I. Zahed, ) Phys. Rev.{\bf C88}, 044915 (2013). 
\bibitem{ref49} S. Chatrchyan  et al., CMS Collaboration, Phys. Lett. {\bf B718}, 795 (2013). 
\bibitem{ref50} G. Aad  et al., ATLAS Collaboration, Phys. Rev. Lett. {\bf 110}, 182302 (2013). 
\bibitem{ref51} S. Chatrchyan  et al., CMS Collaboration, arXiv:hep-ex/1307.3442(2013). 
\bibitem{ref52} The Durham HepData Project, http://hepdata.cedar.ac.uk/view/ins1123117.
\bibitem{ref53} The Durham HepData Project, http://hepdata.cedar.ac.uk/view/ins890166.
\bibitem{ref54} CERN/ROOT/v123.

\end{thebibliography}
\end{document}